\documentclass[a4paper,titlepage,fontsize=11pt,DIV=12,headings=normal]{scrartcl}
\usepackage[affil-it]{authblk}
\usepackage{moreverb}

\usepackage[colorlinks,bookmarksopen,bookmarksnumbered,citecolor=red,urlcolor=red]{hyperref}

\newcommand\BibTeX{{\rmfamily B\kern-.05em \textsc{i\kern-.025em b}\kern-.08em
T\kern-.1667em\lower.7ex\hbox{E}\kern-.125emX}}

\usepackage{url}
\usepackage{epsfig}
\usepackage{amssymb}
\usepackage{amsmath}
\usepackage{amsfonts}

\usepackage{xargs}                      
\usepackage[pdftex,dvipsnames]{xcolor}  

\usepackage{adjustbox}
\usepackage{array}
\usepackage{booktabs}
\usepackage{multirow}

\newcolumntype{R}[2]{%
    >{\adjustbox{angle=#1,lap=\width-(#2)}\bgroup}%
    l%
    <{\egroup}%
}

\begin{document}

\huge
\begin{centering}
Why Johnny Can't Use Stego: a Human-oriented Perspective on the Application of Steganography

~

\normalsize

\large
Steffen Wendzel$^{1,2}$\\

\normalsize
$^1$ Fraunhofer FKIE, Germany\\
$^2$ Worms University of Applied Sciences, Germany\\

~
\end{centering}


\normalsize
\textbf{Abstract.}
Steganography is the discipline that deals with concealing the existence of secret communications. Existing research already provided several fundamentals for defining steganography and presented a multitude of hiding methods and countermeasures for this research discipline.

\textit{Contribution.}
We identified that no work exists that discusses the process of applying steganography from an individual's perspective. This paper presents a phase model that explains pre-conditions of applying steganography as well as the decision-making process and the final termination of a steganographic communication. The model can be used to explain whether an individual \emph{can} use steganography and to explain whether and why an individual \emph{desires} to use steganography. Moreover, the model can be used in research publications to indicate the addressed model's phase of scientific contributions. Furthermore, our model can be used to teach the process of steganography-application to students.

~


\section{Introduction}

\emph{Steganography} deals with the concealing of messages. In informal words, steganography can be defined as \emph{the practice of undetectably communicating a message in a cover object}~\cite{fridrich2009steganography}, i.e. within a carrier. For instance, one steganographic method represents a secret message in written notes of a music composition~\cite{petitcolas1999information}.

The goal of steganography, in comparison to cryptography, is not to hide the content of a secret message but to hide the existence of a secret message~\cite{Anderson:2006:LS:2312088.2312781}. For this reason, steganography is applied in domains where a hidden transfer of messages is beneficial, such as to circumvent Internet censorship, to transfer secret information between spies and in the military context, to hide illegal data or to perform stealthy malware communications~\cite{fridrich2009steganography,petitcolas1999information,katzenbeisser2016information,zander2007survey,NetStegBook}.
Hidden information can be transferred in analog formats, such as in paper letters, but is primarily embedded into digital media, such image, audio and video files~\cite{fridrich2009steganography,petitcolas1999information,katzenbeisser2016information}. In addition, the transfer of hidden information over networks became popular in recent years~\cite{NetStegBook,Zielinska:2014:TS:2566590.2566610}. Another recent trend in the domain is the embedding of secret information into cyber-physical systems~\cite{Wendzel:3SL,CPS:InformationFlowProtection,CCAttacksPervasiveComputing}.

\paragraph*{Addressed Problem}
Several hundred works on Steganography have been published in the last few years. Most of these publications either present new or improved hiding methods or they present new or improved methods to counter these hiding methods.

In contrast, our recent research addresses fundamentals of this discipline, such as the unified and comparable description of hiding methods~\cite{UnifiedDescrMethod} or the evaluation of the novelty and applicability of hiding methods~\cite{Wendzel:jucs:creativity}. However, we discovered that another fundamental aspect of steganography requires a clearer analysis, namely the process of applying steganography from a human-oriented perspective.

\paragraph*{Contribution}
For this reason, we introduce a \textbf{phase model that provides a human-oriented perspective on the application of steganography}. It is the first model that describes all necessary steps an individual has to perform to establish and finalize a covert communication. Our model can be used to (1) explain why an individual can or cannot use (or want to use) steganography, (2) as a tool for researchers to clarify which phases of the model are targetted by their work, and (3) to teach the process of steganography-application decision-making to students.

\paragraph*{Overview}
This paper is structured as follows. Section~\ref{Sect:RelWork} presents related work.
We introduce our phase model in Section~\ref{Sect:Model} and discuss it in Section~\ref{Sect:Discussion}. A conclusion and a brief outlook on future work are provided in Section~\ref{Sect:Concl}.

\section{Related Work}\label{Sect:RelWork}

This section highlights related work and distinguishes our work from these existing publications.

An initial terminology for Information Hiding, the discipline of which steganography is considered a sub-discipline, was collected by Pfitzmann in~\cite{Pfitzmann:1996:IHT:647594.731530}. Several publications discuss the origins of steganography, e.g.~\cite{fridrich2009steganography,petitcolas1999information,katzenbeisser2016information,zander2007survey,NetStegBook,Kahn1996,Zielinska:2014:TS:2566590.2566610}. These publications mention several use-cases for steganography-application and, for this reason, shed light on the possible motivations for applying steganography. However, these works do not add a view on the human-oriented process for applying steganography behind these use-cases.

Simmons introduced his so-called \emph{Prisoner's Problem} that provides the fundamental scenario in which steganography is applied~\cite{Simmons1984}. In his scenario, two prisoners try to escape jail. For a successful escape, they need to work together. However, they cannot directly exchange messages. Instead, the only way to exchange messages is to hand over all messages to a so-called \emph{warden} that can read and modify the messages. The prisoners need to apply steganography so that they can plan their escape without letting the warden notice. Simmons describes a generalized case of steganography-application that comprises two important elements that we discuss in our work: the reason that leads to the application of steganography as well as the decision-making process of the prisoners. Modifications of the Prisoner's Problem exist, e.g.~\cite{OutOfBandCCSurvey}.

In own previous work, we provide a unified description system for steganography hiding methods~\cite{UnifiedDescrMethod}. When applied by authors, our description system allows the easier comparison of hiding methods; it also allows to easily spot what work is lacking for a particular hiding method, e.g. a capacity estimation or the evaluation of certain types of countermeasures.
We also developed a framework for the evaluation of a hiding method's creativity~\cite{Wendzel:jucs:creativity}.\footnote{In this sense, \emph{creativity} is a term from Psychology research and defined in terms of novelty and applicability of a scientific invention~\cite{CreativityInScience}.} However, both works address fundamentals of the discipline but like the abovementioned publications do not address human aspects.

Zseby \emph{et al.} provide a testbed designed to teach network covert channels to university students~\cite{Zseby:NetStegLab}. In own previous work \cite{CCSPoster}, we designed an educational communication protocol tailored for teaching the embedding of covert channels into network protocols. Although these publications address a human aspect (education), they focus on teaching technical details and not on the individual's process that leads to the application of steganography.

Brandstetter \emph{et al.} describe results of a research project (\emph{StegIT}) in a book publication~\cite{Brandstetter:StegITbook}. Their project analyses steganography primarily from a social science perspective. Smaldino \emph{et al.} analyse the evolution of cooperation via covert signalling \cite{CoopCovertSignalling}. In their work, covert signalling is a social covert communication, e.g.\ using jokes or ambiguities in a communication. The process of applying steganography as we present it in this work was not discussed in these two publications.

Furthermore, usable security has been heavily addressed in the last ten years, and several publications stated that increasing usable security is crucial or proposed design goals for usable security, e.g.\ cf.~\cite{Lampson:UsableSecurity:HowToGetIt,UsableSecFiveLessons}. However, researching usable security focuses primarily on analysing and increasing the usability of information technology. It does not automatically involve research on a decision-making process or a description of a process that leads to the application of a particular technology as in case of our paper.

For the domains of web browsing, utilization of authentication methods, or the application of cryptography methods, user studies were performed, e.g.\ for PGP in~\cite{WhyJohnnyCantEncrypt,sheng2006johnny}. These publications address a human aspect but they do not provide a model that addresses the decision-making process and rather analyse the security of a particular encryption tool. Moreover, these publications are not focusing on steganography or its tools.

\section{A Human-oriented Model for Steganography-Application}\label{Sect:Model}

In this section, we introduce a human-oriented perspective that describes the phases of applying steganography from the view of an individual. In our model, steganography is either applied for self-use (e.g. hiding data in a local computer's steganographic filesystem) or to covertly communicate with a peer.\footnote{Here, the term `peer' is referring to any subject desired to be part of a covert communication by the individual.}

\subsection{Design Principles}

Our model was designed by following certain principles to support its acceptance within the scientific community as well as for supporting its use in higher education. In particular, the following design principles were applied:

\newpage
\begin{itemize}
 \item \textbf{Focus on the Human-oriented Application of Steganography.} Instead of developing a model with several layers (e.g.\ an interface layer describing a software interface of tools, a layer for the user, or a technical layer for the functioning of the steganographic methods), the goal was develop a single-layered human-oriented model only for describing the process required to apply steganography.
 
 \item \textbf{Simplicity} The model kept as untangled as possible to prevent unnecessary complexity which would not benefit its acceptance or clarity.
 
 \item \textbf{Cross-domain Applicability.} Our model is not created with only one steganographic domain, e.g.\ digital media steganography, in mind. Instead, it should be applicable to all domains, from linguistic steganography to cyber-physical systems steganography.
\end{itemize}

\subsection{Model Description and Illustration}

\begin{figure}[h!]
  \begin{centering}
  \includegraphics[width=0.8\textwidth]{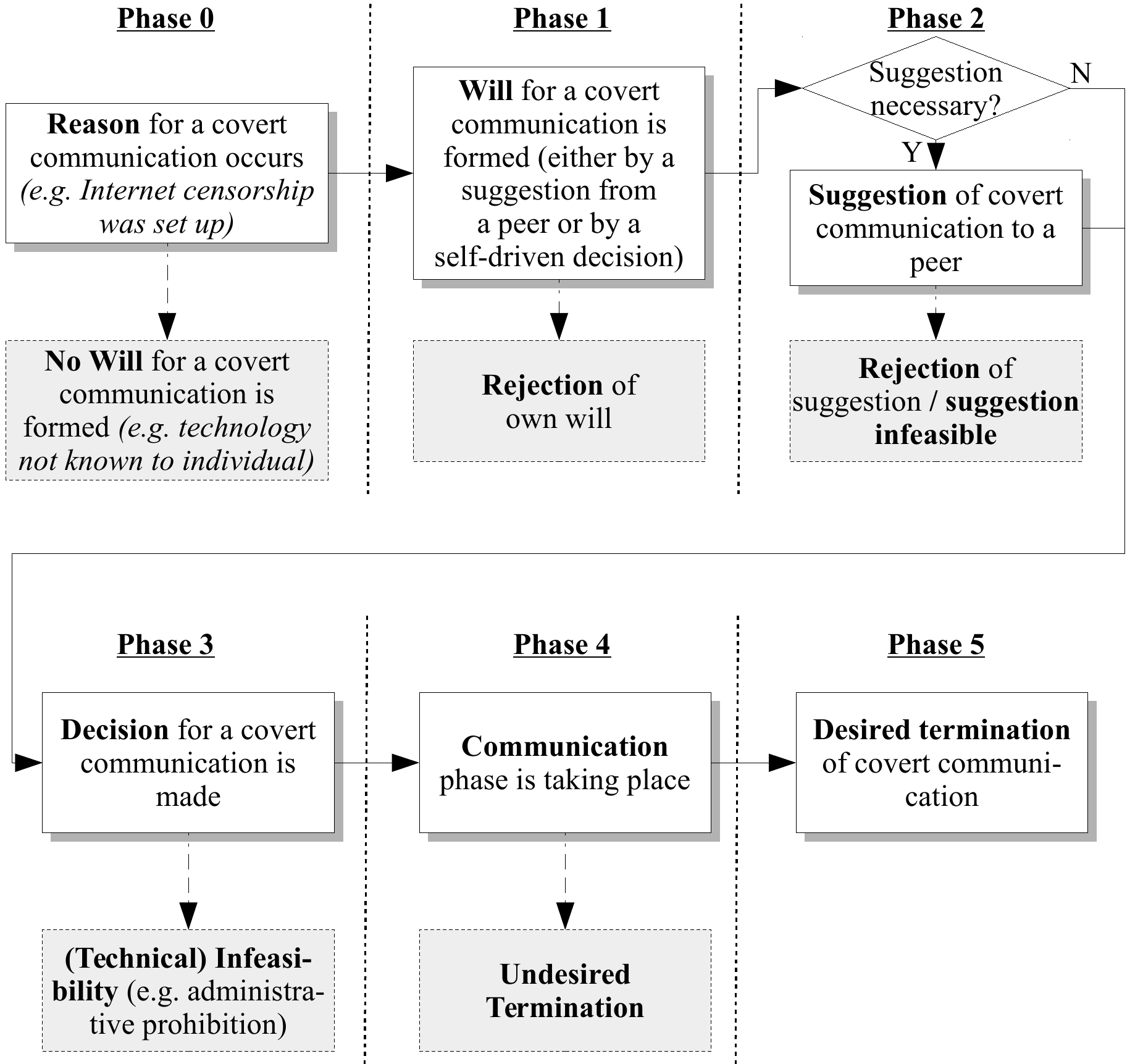}
  \caption{\textit{Covert Communication Phase Model.}
	This model shows the phases that an individual has to go through to prepare, perform,
	and finally terminate a covert communication.\label{fig:PhaseModel:Human}}
	\end{centering}
      \end{figure}

We will now describe all phases of our model. The whole model is split into six phases (0--5) and is visualized in Figure~\ref{fig:PhaseModel:Human}. The process starts at the upper left side with phase 0.

\subsubsection*{Phase 0: Occurrence of a Reason} This is a pre-phase to enter the phase model. Entering the phase is triggered by an external event or by an intrinsic motivation that provides a \emph{reason} for a steganographic communication. In Simmons' Prisoner's Problem, the reason is the fact that the prisoners are locked in jails and possess the desire to escape in a joint-work process.

No individual will apply steganography without an initial reason for its use. Such a reason can, for instance, be the occurrence of Internet censorship so that two parties are not able to exchange their political opinions any further. Another reason for the use of steganography would be to teach steganography to students in higher education. Several additional scenarios for the application of steganography can be found in~\cite{fridrich2009steganography,petitcolas1999information,katzenbeisser2016information,zander2007survey,NetStegBook,Zielinska:2014:TS:2566590.2566610}.

Phase 0 can be reached unknowingly by an individual, e.g.\ if messages are observed by an adversary without the knowledge of the individual. A similar reason for reaching phase 0 unknowingly would be if an adversary unnoticeably manipulates the communication between an individual and a peer via a man-in-the-middle attack, ensuring that only filtered content reaches the peer.

\paragraph*{Reasons not to Reach Phase 0:} Obviously, the only reason not to reach phase 0 is the lack of a reason.

\paragraph*{Illustration:} Alice and Bob live in different places of the same country and only meet every two or three weeks in person for discussing about the government's handling of human rights. However, they perform a regular online communication to intensify their exchange. Recently, both experienced that e-mails sent by Alice to Bob are not arriving in Bob's inbox (and vice versa). After an analysis they discover that only e-mails without illicit opinions arrive in each other's inbox. Alice and Bob suspect that a government organization may actively block their e-mails.

\subsubsection*{Phase 1: Formation of a Will} An individual reaches phase 1 when she forms the will for performing a steganographic communication.
A will is formed either in a self-driven manner, i.e. if the individual possesses the necessary know-how about steganography, or if the use of steganography is suggested by a peer and the individual starts to consider to use it. In general, a will can only be formed after an initial interest in steganography was created. However, interest in something must first be triggered and then maintained before it can emerge and eventually become well-developed (the related process is described in a four-phase model by Hidi and Renninger~\cite{FourPhaseInterestModel}).

\paragraph*{Reasons not to Reach Phase 1:}
An individual does \emph{not} reach phase 1 if there is a reason but no will for using steganography. Several reasons are imaginable for an individual not to form a will to use steganography, including but not limited to the following reasons:
\begin{itemize}
\item unconsciousness about the steganography-application reason and thus the fact of being phase 0 at all;
\item acceptance of the fact that a communication may be observed or manipulated (acceptance of the reason that caused phase 0);
\item lack of interest in steganography, i.e. the interest was maybe not triggered or was not maintained, so that it could not develop~\cite{FourPhaseInterestModel};
\item similarly to a lack of interest, the individual possibly has a lack of motivation for using steganography, i.e.\ security is a less important concern in comparison to a functioning communication for the individual (a similar reason was provided in other research for not using the PGP encryption software~\cite{WhyJohnnyCantEncrypt});
\item lack of knowledge about the existence of steganography itself;
\item consideration of steganography as inappropriate (e.g. rejection of a protection that relies on any form of \emph{security by obscurity}~\cite{petitcolas1999information,mercuri2003security}\footnote{Due to the introduction of steganographic keys~\cite{Anderson1996,Anderson:2006:LS:2312088.2312781,provos2003hide}, Kerckhoffs' Principle of security only relying on a key can be applied to steganography as well. However, this fact is not necessarily known to an individual.});
\item negative perception of the value of steganography (in the context of the reason from phase 0);\footnote{For instance, Devaraj \emph{et al.} have shown that the value of anonymity, which is another discipline of Information Hiding~\cite{petitcolas1999information}, differs significantly depending on several social and cultural aspects such as gender, race, education and income level~\cite{Devaraj2015}. This may be applicable to steganography, too.}
\item lack of understanding and lack of experience with steganography as a technology -- this may be linked to a lack of feedback as was pointed out for the case of the PGP encryption software~\cite{WhyJohnnyCantEncrypt};
\item fear of detection when a known steganography method is used;
\item lack of trust into the potential peer~\cite{Brandstetter:StegITbook}{(Chapter 1)};
\item lack of similarity between the individual and her peer(s): Smaldino \emph{et al.} state that pairs of similar individuals can more effectively coordinate and understand their covert signalling; however, this aspect is limited to non-technical covert communication~\cite{CoopCovertSignalling} and would thus mostly apply to linguistic steganography;
\item use of an alternative technology (e.g. \emph{Tor}-based onion routing~\cite{dingledine2004tor}) instead of steganography caused by the reason in phase 0.
\end{itemize}

\paragraph*{Illustration:} Fearing an observation, Alice considers to perform a communication that is not solely encrypted but hidden from an observer. She forms the will to use steganography for her communication with Bob.

\subsubsection*{Phase 2: Suggestion to a Peer} In most cases, at least one peer is required for the individual to exchange of covert information with. For this reason, in phase 2, the individual suggests the application of steganography to a peer. This is also the first phase where a coordination with a peer is a necessity.

\paragraph*{Reasons not to Reach Phase 2:}
A reason not to reach phase 2 is if a peer is required for the steganographic communication but the individual rejected her own will to suggest the use of steganography. A possible reason is the awareness that (one of) the desired peers(s) would be unable to use steganography.

Phase 2 is also not reached if the application of steganography was suggested by a peer to the individual (phase 1) and no other peers are required to be suggested to use steganography.

Another reason is the case when steganography is applied by the individual to store and retrieve covert data only with herself. For instance, filesystem steganography~\cite{anderson1998steganographic,mcdonald1999stegfs} may be used on a local computer by only one individual to store and access own secret data. However, a single individual's will is only sufficient for applying steganography if the individual is both the sender and receiver of all covert information exchange caused by the reason of phase 0.

In such cases where no (additional) peer is required, phase 2 is not considered and phase 1 directly leads to phase 3.

\paragraph*{Illustration:} At their next physical meeting, Alice suggests Bob to use steganography for their communication.

\subsubsection*{Phase 3: Decision to use Steganography} Phase 3 is reached if all necessary participants decided to apply steganography for their communication.

In Simmons' Prisoner's Problem, phases 1--3 are merged in brief statements (the prisoners want to perform a covert communication and they \emph{accept} the condition to exchange messages over the warden). However, the process of suggesting the use of steganography to each prisoner's peer is missing and the formation of the will is not discussed but just made.

\paragraph*{Reasons not to Reach Phase 3:} This phase is not reached if one necessary peer rejected the suggested use of steganography. The phase is also not reached if, e.g.\ due to a blocked network communication or the impossibility of a physical meeting, no communication channel can be used to suggest the use of steganography (i.e., step 2 failed).

\paragraph*{Illustration:} In the previous phase, Alice suggested Bob to use steganography. However, Bob never heard of steganography. He is sceptic and prefers to use PGP instead. However, Alice convinces Bob to use steganography for their communication as this could prevent to raise suspicion at the observer. They decide to additionally encrypt all secret messages before they are embedded into a steganographic carrier.

\subsubsection*{Phase 4: Communication (Steganography Application)} Finally, the participants perform their covert communication. They first initiate their covert communication. This is achieved by selecting a \emph{steganographic system}~\cite{westfeld1999attacks}. Therefore, they either buy or download one of the steganography software or hardware tools (e.g.\ those listed in \cite{Wang:2004:CWS:1022594.1022597} or that can be found via an online search) from the Internet or a store. Alternatively, they can implement an own steganography system. They eventually exchange steganographic keys and finally use their steganographic system. The general process of using steganography is exemplified with image steganography in \cite{zollner1998modeling}. The communication phase happens within a spectrum reaching from simple steganography, e.g. LSB manipulation in a single image file~\cite{johnson2000survey} or simple forms of linguistic steganography~\cite{petitcolas1999information}, to state-of-the-art methods, e.g. adaptive steganography methods for network steganography~\cite{NetStegBook} or advanced methods for digital media steganography. The technical options and their details for performing steganographic communications are well addressed by several of the already cited references. However, as we highlight a human-oriented aspect, technical details are not in our focus.

\paragraph*{Reasons not to Reach Phase 4:}
Phase 4 is not reached if due to (technical) reasons, a covert communication cannot be established. Possible reasons are the presence of a warden that modifies or blocks the carrier (e.g. digital images~\cite{chandramouli2003image} or network packets~\cite{fisk2002eliminating}) on the communication path between the participants. In this case, the individual will either stop using her steganographic system or will re-try using the steganographic system at a later time. Alternatively, the individual and the peers may select another steganographic system or switch to an alternative anti-censorship technology.

\paragraph*{Illustration:} During their meeting, Alice and Bob decide to use three different tools for steganography: two tools utilize digital images and one hides secret data in network transmissions. To test, whether one of the three tools will be blocked, they send and reply one message with each tool -- similar to a TCP three-way-handshake.  Back at home, Alice sends an e-mail with two photos generated by tools \#1 and \#2 to Bob and, in addition, she transfers steganographic network traffic to him with tool \#3. Bob receives the messages and answers on the following day with the same three methods. Alice receives all three acknowledgement messages and confirms the receipt of all the messages with a single message using one of the three tools. From now on, Alice and Bob use only one tool at a time, knowing that it will be likely that all three tools will pass the observation.

\subsubsection*{Phase 5: Desired Termination} Finally, after the participants finished their exchange of secret data, they terminate their covert communication. Several steganographic protocols describe how connections (sessions) are (initiated and) terminated in a technical way, e.g.\ using simple start and stop flags~\cite{ray2008protocol}. From the individual's perspective, however, this means to end the current utilization of a steganographic system (e.g.\ in form of a tool). Optionally, the individual can remove the related tool that implements the steganographic system as well as the related digital files or e-mails from her computer or from a server; she can moreover destroy hardware components and any type of evidence that would point an adversary to the fact that she used steganography. The elimination of evidence could be represented in a `Phase 6' but does not appear necessary as it is an optional task.

\paragraph*{Reasons not to Reach Phase 5:}
This desired termination phase is not reached if the communication was terminated in an undesired way, e.g. due to network problems or because an adversary/observer terminated the communication. The handling of the situation would be the same as in case of the desired termination (e.g.\ optional removal of the tool, related digital files, and destruction of evidence for the use of steganography).

\paragraph*{Illustration:} For a summer break, Bob will be out of the country. Alice and Bob decide to stop their steganographic communication and may start a new discussion in autumn.

\section{Discussion}\label{Sect:Discussion}

After the previous section introduced our model, we now discuss certain additional aspects of it, including potential drawbacks.

\subsubsection*{General Discussion}

When we discuss why an individual cannot or does not want to use steganography, a fundamental aspect should be mentioned: although knowledge on steganography was kept secret for a long time, textbooks and other publications are publicly available since decades. Publications and web-based discussions on steganography moreover increase continuously since years.\footnote{A keyword search for the term `steganography' on Google Scholar revealed an average yearly increase of 26.75\% of publications that match the keyword for the years 1995-2015.}
For this reason, we cannot assume that steganography is not widely used in public due to a lack of `early adopters' as defined by~\cite{rogers2010diffusion}. Instead of being a mainstream technology, steganography is seen as a niche technology used only for special circumstances. In our model, those circumstances provide the \emph{reasons} to enter phase 0.

In addition, the increase of scientific publications during the last decades does not necessarily influence the extend in which steganography is used because there can be large time-gaps between the publication of scientific ideas (and even re-inventions of the same) and their actual application \cite{CreativityInScience}.
However, our model shows several reasons that can prevent the application of steganography by an individual. Especially a broad range of reasons already influences whether phase 1 will be reached.

In our model, the duration of a phase is not defined; it can reach from seconds to years. Each phase can be exited either based on an own decision (except for phase 0) or due to external circumstances (e.g.\ a blocked communication channel that leads to an undesired termination of a covert communication).

Our model contains several conditions that describe when a certain phase is not reached. The individual, at these points, can either stay in the previous phase or can go back to any earlier phase, including phase 0. For exiting phase 0, two scenarios must be distinguished:
If phase 0 was entered due to an external reason, maybe even unknowingly for the individual (e.g.\ due to unnoticed surveillance), it is not foreseen that the individual can exit the model own her own will unless the reason for phase 0 is not present anymore. If, on the other hand, the reason for entering phase 0 was self-defined, e.g.\ for the self-defined purpose of teaching steganography in a class, the individual can eliminate this reason at any time, thus exit the model.

Phase 0 may be enforced for literally every Internet user by default as every user could face surveillance. As it would be impossible to define a threshold to enter phase 0 (the exact extend of surveillance may not be known or measurable at all), researchers working with the model should define a plausible reason to enter phase 0 in dependence of an individual's circumstances. For instance, more reasons would be plausibly linkable to a phase 0 for a government-criticizing journalist than for most other citizens.

\subsubsection*{Potential Drawbacks}
It is important to emphasize that our model is not designed to be static. Instead, it can be modified by any researcher in new research publications if options for the model's improvement are identified.

Although our proposed model was conducted in a meticulous process, it is not necessarily complete. It is imaginable that we did not consider a special or rare scenario for our model, especially as one design principle was to keep the model untangled. Such a special case is possibly integratable into the model -- either in an existing phase of the model or by introducing a new phase. For instance, the decision of not having a `phase 6' -- as pointed out in the description of phase 5 -- was to keep the model compact. However, one might argue that such a `phase 6' would be beneficial due to reasons which we did not consider.

Moreover, our model does not consider detailed psychological aspects and it does not focus on the usability of steganographic tools. Following our design goals, these aspects were not intended to be part of this work.

\section{Conclusion and Future Work}\label{Sect:Concl}

We present a phase model that provides a human-oriented perspective on the application of steganography. Our phase model adds several aspects to the discussion of applying steganography that were not considered in detail before. Starting with the provision of a reason for a covert communication, the model leads through all necessary steps in a structured manner until a desired termination of the covert communication is reached. Scientific publications can use our model to tell which phases they consider in their work and to analyse whether additional phases could be considered as well. In addition, the model can serve as a tool to teach the process of steganography-application to students.

Future work may comprise to analyse the usability of steganography tools and to determine design rules for their improvement. Such a research could be conducted in similar ways as it was done for cryptographic software, authentication concepts or web-based security tools.

\bibliographystyle{abbrv}
\bibliography{bmc_article}      

\begin{thebibliography}{10}

\bibitem{Anderson1996}
R.~Anderson.
\newblock Stretching the limits of steganography.
\newblock In {\em Information Hiding: First International Workshop Cambridge,
  U.K., May 30 -- June 1, 1996 Proceedings}, pages 39--48, Berlin, Heidelberg,
  1996. Springer Berlin Heidelberg.

\bibitem{anderson1998steganographic}
R.~Anderson, R.~Needham, and A.~Shamir.
\newblock The steganographic file system.
\newblock In {\em International Workshop on Information Hiding}, pages 73--82.
  Springer, 1998.

\bibitem{Anderson:2006:LS:2312088.2312781}
R.~J. Anderson and F.~A. Petitcolas.
\newblock On the limits of steganography.
\newblock {\em IEEE Journal on Selected Areas in Communications},
  16(4):474--481, Sept. 2006.

\bibitem{UsableSecFiveLessons}
D.~Balfanz, G.~Durfee, R.~E. Grinter, and D.~K. Smetters.
\newblock In search of usable security: Five lessons from the field.
\newblock {\em IEEE Security \& Privacy Magazine}, 2(5):19--24, 2004.

\bibitem{Brandstetter:StegITbook}
M.~Brandstetter, M.~Schmidberger, and S.~Sommer, editors.
\newblock {\em Die Funktion ``verdeckter Kommunikation''. Impulse für eine
  Technikfolgenabschätzung zur Steganographie}.
\newblock Number~9 in Soziale Arbeit -- Social Issues. LIT Verlag,
  Vienna/Berlin, 2010.
\newblock in German.

\bibitem{OutOfBandCCSurvey}
B.~Carrara and C.~Adams.
\newblock Out-of-band covert channels -- a survey.
\newblock {\em ACM Comput. Surv.}, 49(2, article 23):1--36, June 2016.

\bibitem{chandramouli2003image}
R.~Chandramouli, M.~Kharrazi, and N.~Memon.
\newblock Image steganography and steganalysis: Concepts and practice.
\newblock In {\em International Workshop on Digital Watermarking}, pages
  35--49. Springer, 2003.

\bibitem{Devaraj2015}
S.~Devaraj, M.~Alfred, K.~C. Madathil, and A.~K. Gramopadhye.
\newblock {\em An Investigation of the Factors that Predict an {Internet}
  User's Perception of Anonymity on the Web}, pages 311--322.
\newblock Springer International Publishing, Cham, 2015.

\bibitem{dingledine2004tor}
R.~Dingledine, N.~Mathewson, and P.~Syverson.
\newblock Tor: The second-generation onion router.
\newblock Technical report, DTIC Document, 2004.
\newblock
  \url{http://oai.dtic.mil/oai/oai?verb=getRecord\&metadataPrefix=html\&identifier=ADA465464}.

\bibitem{fisk2002eliminating}
G.~Fisk, M.~Fisk, C.~Papadopoulos, and J.~Neil.
\newblock Eliminating steganography in {Internet} traffic with active wardens.
\newblock In {\em International Workshop on Information Hiding}, pages 18--35.
  Springer, 2002.

\bibitem{fridrich2009steganography}
J.~Fridrich.
\newblock {\em Steganography in digital media: principles, algorithms, and
  applications}.
\newblock Cambridge University Press, New York, NY, USA, 2009.

\bibitem{FourPhaseInterestModel}
S.~Hidi and K.~A. Renninger.
\newblock The four-phase model of interest development.
\newblock {\em Educational Psychologist}, 41(2):111--127, 2006.

\bibitem{CPS:InformationFlowProtection}
G.~Howser.
\newblock Using information flow methods to secure cyber-physical systems.
\newblock In {\em Critical Infrastructure Protection IX}, volume 466 of {\em
  IFIP AICT}, pages 185--205, Cham, 2015. Springer.

\bibitem{johnson2000survey}
N.~F. Johnson and S.~Katzenbeisser.
\newblock A survey of steganographic techniques.
\newblock In {\em Information hiding}, pages 43--78. Norwood, MA: Artech House,
  2000.

\bibitem{Kahn1996}
D.~Kahn.
\newblock The history of steganography.
\newblock In R.~Anderson, editor, {\em Information Hiding: First International
  Workshop Cambridge, U.K., May 30 -- June 1, 1996 Proceedings}, pages 1--5,
  Berlin, Heidelberg, 1996. Springer Berlin Heidelberg.

\bibitem{katzenbeisser2016information}
S.~Katzenbeisser and F.~Petitcolas.
\newblock {\em Information hiding}.
\newblock Artech House, Norwood, MA, 2016.

\bibitem{Lampson:UsableSecurity:HowToGetIt}
B.~Lampson.
\newblock Usable security: How to get it.
\newblock {\em Communications of the ACM}, 52(11):25--27, 2009.

\bibitem{NetStegBook}
W.~Mazurczyk, S.~Wendzel, S.~Zander, A.~Houmansadr, and K.~Szczypiorski.
\newblock {\em Information Hiding in Communication Networks: Fundamentals,
  Mechanisms, and Applications}.
\newblock IEEE Press Series on Information and Communication Networks Security.
  Wiley-IEEE Press, Hoboken, NJ, 2016.

\bibitem{mcdonald1999stegfs}
A.~D. McDonald and M.~G. Kuhn.
\newblock Stegfs: A steganographic file system for {Linux}.
\newblock In {\em International Workshop on Information Hiding}, pages
  463--477. Springer, 1999.

\bibitem{mercuri2003security}
R.~T. Mercuri and P.~G. Neumann.
\newblock Security by obscurity.
\newblock {\em Communications of the ACM}, 46(11):160, 2003.

\bibitem{petitcolas1999information}
F.~A. Petitcolas, R.~J. Anderson, and M.~G. Kuhn.
\newblock Information hiding-a survey.
\newblock {\em Proceedings of the IEEE}, 87(7):1062--1078, 1999.

\bibitem{Pfitzmann:1996:IHT:647594.731530}
B.~Pfitzmann.
\newblock Information hiding terminology - results of an informal plenary
  meeting and additional proposals.
\newblock In {\em Proceedings of the First International Workshop on
  Information Hiding}, pages 347--350, London, UK, 1996. Springer-Verlag.

\bibitem{provos2003hide}
N.~Provos and P.~Honeyman.
\newblock Hide and seek: An introduction to steganography.
\newblock {\em IEEE Security \& Privacy}, 1(3):32--44, 2003.

\bibitem{ray2008protocol}
B.~Ray and S.~Mishra.
\newblock A protocol for building secure and reliable covert channel.
\newblock In {\em Privacy, Security and Trust, 2008. PST'08. Sixth Annual
  Conference on}, pages 246--253. IEEE, 2008.

\bibitem{rogers2010diffusion}
E.~M. Rogers.
\newblock {\em Diffusion of innovations}.
\newblock Simon and Schuster, New York, NY, 2010.

\bibitem{sheng2006johnny}
S.~Sheng and L.~Broderick.
\newblock Why {Johnny} still can’t encrypt: evaluating the usability of email
  encryption software.
\newblock In {\em Symposium On Usable Privacy and Security}, 2006.
\newblock
  \url{https://cups.cs.cmu.edu/soups/2006/posters/sheng-poster\_abstract.pdf}.

\bibitem{Simmons1984}
G.~J. Simmons.
\newblock {\em The Prisoners' Problem and the Subliminal Channel}, pages
  51--67.
\newblock Springer US, Boston, MA, 1984.

\bibitem{CreativityInScience}
D.~K. Simonton.
\newblock {\em Creativity in Science. Chance, Logic, Genius and Zeitgeist}.
\newblock Cambridge University Press, Cambridge, 2004.

\bibitem{CoopCovertSignalling}
P.~E. Smaldino, T.~J. Flamson, and R.~McElreath.
\newblock Evolution of cooperation via covert signaling.
\newblock {\em arXiv preprint}, 1511.04788, 2015.
\newblock \url{http://arxiv.org/abs/1511.04788}.

\bibitem{CCAttacksPervasiveComputing}
N.~Tuptuk and S.~Hailes.
\newblock Covert channel attacks in pervasive computing.
\newblock In {\em Proc. 2015 IEEE International Conference on Pervasive
  Computing and Communications (PerCom)}, pages 236--242, Piscataway, NJ, USA,
  2015. IEEE.

\bibitem{Wang:2004:CWS:1022594.1022597}
H.~Wang and S.~Wang.
\newblock Cyber warfare: Steganography vs. steganalysis.
\newblock {\em Commun. ACM}, 47(10):76--82, Oct. 2004.

\bibitem{Wendzel:3SL}
S.~Wendzel, B.~Kahler, and T.~Rist.
\newblock Covert channels and their prevention in building automation
  protocols: A prototype exemplified using {BACnet}.
\newblock In {\em Green Computing and Communications (GreenCom), 2012 IEEE
  International Conference on}, pages 731--736, Piscataway, NJ, USA, November
  2012. IEEE.

\bibitem{CCSPoster}
S.~Wendzel and W.~Mazurczyk.
\newblock Poster: An educational network protocol for covert channel analysis
  using patterns.
\newblock In {\em 23rd ACM Conference on Computer and Communications Security
  (CCS'16)}, New York, NY, 2016. ACM.
\newblock in Press.

\bibitem{UnifiedDescrMethod}
S.~Wendzel, W.~Mazurczyk, and S.~Zander.
\newblock Unified description for network information hiding methods.
\newblock {\em arXiv preprint}, 1512.07438, 2015.
\newblock \url{http://arxiv.org/abs/1512.07438}, currently under review.

\bibitem{Wendzel:jucs:creativity}
S.~Wendzel and C.~Palmer.
\newblock Creativity in mind: Evaluating and maintaining advances in network
  steganographic research.
\newblock {\em Journal of Universal Computer Science (J.UCS)},
  21(12):1684--1705, 2015.
\newblock
  \url{http://www.jucs.org/jucs\_21\_12/creativity\_in\_mind\_evaluating}.

\bibitem{westfeld1999attacks}
A.~Westfeld and A.~Pfitzmann.
\newblock Attacks on steganographic systems.
\newblock In {\em International workshop on information hiding}, pages 61--76.
  Springer, 1999.

\bibitem{WhyJohnnyCantEncrypt}
A.~Whitten and J.~D. Tygar.
\newblock Why {Johnny} can't encrypt: A usability evaluation of {PGP} 5.0.
\newblock In {\em Proceedings of the 8th Conference on USENIX Security
  Symposium - Volume 8}, SSYM'99, pages 14--14, Berkeley, CA, USA, 1999. USENIX
  Association.

\bibitem{zander2007survey}
S.~Zander, G.~Armitage, and P.~Branch.
\newblock A survey of covert channels and countermeasures in computer network
  protocols.
\newblock {\em IEEE Communications Surveys \& Tutorials}, 9(3):44--57, 2007.

\bibitem{Zielinska:2014:TS:2566590.2566610}
E.~Zieli\'{n}ska, W.~Mazurczyk, and K.~Szczypiorski.
\newblock Trends in steganography.
\newblock {\em Commun. ACM}, 57(3):86--95, Mar. 2014.

\bibitem{zollner1998modeling}
J.~Z{\"o}llner, H.~Federrath, H.~Klimant, A.~Pfitzmann, R.~Piotraschke,
  A.~Westfeld, G.~Wicke, and G.~Wolf.
\newblock Modeling the security of steganographic systems.
\newblock In {\em International Workshop on Information Hiding}, pages
  344--354. Springer, 1998.

\bibitem{Zseby:NetStegLab}
T.~Zseby, F.~I. Vazquez, V.~Bernhardt, D.~Frkat, and R.~Annessi.
\newblock A network steganography lab on detecting {TCP/IP} covert channels.
\newblock {\em IEEE Transactions on Education}, 59(3):224--232, 2016.

\end{thebibliography}

\end{document}